\documentstyle[pre,aps,multicol]{revtex}
\def\O{{\bf\Omega}}
\def\R{{\bf R}}
\def\D{{\bf D}}
\def\U{{\bf U}}
\def\Si{{\rm Si}}
\def\Ci{{\rm Ci}}
\def\i{{\rm i}}
\def\Pib{{\bf \Pi}_0}
\def\Pie{{\bf \Pi}_1}

\begin{document}
\draft \preprint{}
\title{Spin Dephasing in the Extended Strong Collision Approximation}

\author{Wolfgang R. Bauer*\cite{bylineWolfgang},
Walter Nadler**}

\address{
* Medizinische Universit\"atsklinik,
Universit\"at W\"urzburg, Josef Schneider Strasse 3, D-97080
W\"urzburg, Germany}

\address{
** Fakult\"at f\"ur  Physik, Bergische Universit\"at,
Gau{\ss}str.20, D-42097 Wuppertal, Germany, and \\ NIC c/o
Forschungszentrum J\"ulich, D-52725 J\"ulich, Germany}

\date{\today}

\maketitle

\begin{abstract}
For Markovian dynamics of field fluctuations we present here an
extended strong collision approximation, thereby putting our
previous strong collision approach (Phys. Rev. Lett. 83 (1999)
4215) into a systematic framework. Our new approach provides
expressions for the free induction and spin echo magnetization
decays that may be solved analytically or at least numerically. It
is tested for the generic cases of dephasing due to an
Anderson-Weiss process and due to restricted diffusion in a linear
field gradient.
\end{abstract}

\begin{multicols}{2}

\narrowtext

\section{Introduction}

The understanding of spin dephasing is of paramount interest in
all fields of nuclear magnetic resonance (NMR) sciences. In NMR
spectroscopy it determines the line shape, in NMR imaging it is --
besides longitudinal relaxation -- the major mechanism determining
the contrast and contains morphological as well as functional
information.

The processes contributing to spin dephasing are related to the
spin environment. In biological tissues, for example, spin
dephasing may result from dipole-dipole interaction of water
proton spins with paramagnetic ions like Fe$^{2+}$. Another cause
is diffusion within inhomogeneous magnetic fields generated by
native or contrast agent induced susceptibility differences that
are related to tissue composition and/or cellular and sub cellular
compartments. In magnetic resonance imaging spin dephasing in
external gradient fields is exploited to get information about
diffusion within biological systems. These diffusion sensitive
imaging techniques are applied to study tissue anisotropy and
restrictions of diffusion that are given by membranes of cells and
sub cellular structures.

Essential for dephasing of spins are the field fluctuations that
induce the phase modulations. It is important to note that in
biological tissues the relevant processes cover almost the whole
range of time scales. For example, the dynamics of interactions of
water proton spins with paramagnetic macromolecules as ferritin is
so fast that it can be considered to be within the motional
narrowing regime. On the other hand, dephasing of spins in
magnetic field gradients around larger vessels is almost coherent.
i.e. it is in the static dephasing regime \cite{yablonski}. Hence,
for biological applications it is important to obtain results from
theory that are valid over the whole motion regime. However, in
most cases this is not possible analytically.

Therefore, most efforts have focused on limiting cases. The {\it
motional narrowing limit} is well investigated and a number of
analytical results were obtained for it\cite{abragham}. The
characteristic of this limit is that the mean phase shift induced
by a field realization is much smaller than one, i.e. $|\delta
\varphi|=\tau\langle\Delta\omega^2\rangle^{1/2}\ll 1$, where the
correlation time $\tau$ gives the mean duration of some field
realization, and $\langle\Delta\omega^2\rangle$ is the variance of
the inhomogeneous field. The relaxation time is then obtained as
$1/T_2=\tau\;\langle\Delta\omega^2\rangle$. In the other limiting
case, i.e. the {\it static dephasing regime}
($\tau\langle\Delta\omega^2\rangle^{1/2}\gg 1$),
Yablonski\cite{yablonski} derived analytical expressions for
coherent dephasing of spins in inhomogeneous fields around
magnetic centers like cylinders or spheres. Kiselev \cite{posse}
extended Yablonski's static dephasing approach by considering
diffusion of spins within local linear gradients. However, this
approach requires that the diffusion length $l$ during dephasing
is within the linear approximation of the inhomogeneous fields
$\omega(x_0+l)\approx \omega(x_0)+\partial_x\omega(x_0)\; l$. Note
that expansion around the limiting cases by perturbation
approaches leads to divergences in the respective other limits.
Therefore, the intermediate motion regime, i.e. almost everything
between the static dephasing and motional narrowing limit, was in
most situations accessible by simulations only \cite{kennan}.

Recently, we used a strong collision (SC) approach to characterize
spin dephasing in a particular situation: An inhomogeneous field
around regularly arranged parallel cylinders filled with a
paramagnetic substance \cite{bauer1}, a model reflecting the
capillary network of the cardiac muscle. The results agreed well
with simulations \cite{kennan} over the whole dynamic range, and
with experimental data \cite{wacker1,wacker2}.

The basic idea behind the SC approach is to replace the original
generator of the Markov process by a simpler one, the SC
generator, that conserves particular features of the original
process. In particular,  by an appropriate choice of its parameter
the SC process reproduces the correlation time of the field
fluctuations induced by the original Markov process. There are
several advantages of the SC approximation. First, it is is
correct both in the motional narrowing and the static dephasing
limits; thereby also the error in the intermediate regime is
reduced considerably, when compared to perturbation approaches.
And second, it provides a simple expression for the magnetization
decay which may be solved analytically or at least numerically.

However, the drawback of the SC approach was -- up to now -- that
is not part of a systematic approximation to or an expansion of
the original generator. Therefore, it was unclear how results
could be improved beyond the SC approximation. The aim of this
paper is to extend the SC approach and provide a framework for a
systematic approximation.

In the next section we will present a formal description of spin
dephasing that will be the basis for our analytical analysis. In
Sect.~III we will introduce the extended strong collision (ESC)
approximation proper and show how it is used to describe free
induction and spin echo decay. In Sect.~IV we will apply it to two
generic cases: spin dephasing induced by an Anderson-Weiss process
\cite{anderson} and by restricted diffusion in a linear field
gradient. We will close the paper with a summary and a discussion
of our results.

\section{Formal Description of Spin-Dephasing}

We assume that dephasing of transversely polarized nuclear spins
exposed to an external field is induced by randomly fluctuating
magnetic perturbation fields with frequency $\omega_i$, where $i$
is a discrete or continuous variable. The transition dynamics
between two distinct states $i$ and $j$ is that of a stationary
continuous time Markov process described by rates $r_{ji}$ for the
transition $i \to j$. The matrix $\R=(r_{ji})$ as the generator of
the Markov process conserves the probability to find a spin within
one state, i.e. $r_{ii}=-\sum_{j\neq i} r_{ji}$. The eigenvalues
$l$ of $\R$ fulfill the condition $l\le 0$ where $l=0$ corresponds
to the equilibrium probability distribution. To simplify the
notation we denote the normalized left and right eigenvectors of
$\R$ as $\langle l|$ and $|l\rangle$, respectively, with $\langle
l' |l\rangle=\delta_{l' l}$.

The time evolution between $t$ and $t+dt$ of the transverse
magnetization of spins in the state $j$  (in polar notation
$m_j=m_{j\; x}-{\rm i} m_{j\;y}$) results from the linear
superposition of the transition and the precession dynamics, i.e.
$\partial_t m_j(t)= \sum_j r_{ji}\;m_i(t)+{\rm i}\;\omega_j\;
m_j(t)$. The precession within the external field was omitted
since it only induces a constant offset of the frequency which may
be gauged to zero. With the diagonal frequency matrix
$\O=(\delta_{ji}\; \omega_i)$ one obtains for the magnetization $|
m\rangle=(m_j)$
\begin{equation}
\label{GBT}
\partial_t |m(t)\rangle\;=(\R+{\rm i}\O)\;|m(t)\rangle\;,
\end{equation}
which is a generalization of the Bloch Torrey equation
\cite{torrey} originally formulated for diffusing spins, i.e.\
$\R\sim\nabla^2$. In most cases it is reasonable to assume that
the initial magnetization $|m(0)\rangle$ is proportional to the
equilibrium probability distribution $|0\rangle$, e.g. when free
diffusion is considered this would imply a homogeneous transverse
magnetization. Equation~(\ref{GBT}) then provides the time
evolution of the transverse magnetization (free induction decay)
as
\begin{equation}
|m(t)\rangle\;=\exp[(\R+{\rm i}\O)\;t]\;|0\rangle\;, \label{FID1}
\end{equation}
where the initial magnetization was normalized to one. The overall
magnetization is then determined as
\begin{eqnarray}
M(t)&=&\langle 0|m(t)\rangle\cr\cr &=&\langle 0|\exp\left[(\R+{\rm
i}\O)\;t\right]\;|0\rangle\label{Magn}\;.
\end{eqnarray}
The free induction decay as determined by Eqs.~(\ref{FID1}) and
(\ref{Magn}) results from coherent and incoherent spin dephasing.
The incoherent contribution is determined from spin echo
experiments. In-plane polarized spins are rotated by a $180^0$
($\pi$-pulse) after a time $t/2$. This pulse transforms the
original magnetization $|m(t/2)\rangle$ to its complex adjoint
$|m^*(t/2)\rangle=\exp[(\R-{\rm i}\O)\;t/2]\;|0\rangle$. This
procedure cancels the coherent spin dephasing after the time $t$
(echo time), i.e. the decay of magnetization at $t$ is solely due
to incoherent spin dephasing. The time course of the magnetization
after the pulse, i.e. for times $t' > t/2$, is determined by
\begin{equation}
|m(t')\rangle\;=\exp[(\R+{\rm i}\;\O)(t'-t/2)]\exp[(\R-{\rm
i}\O)t/2]|0\rangle \;,
\end{equation}
i.e. the overall spin echo magnetization at the echo time $t$ is
\begin{eqnarray}
M_{SE}(t)&=&\langle 0|\exp[(\R+{\rm i}\;\O)t/2]\;\exp[(\R-{\rm
i}\O)\;t/2]\;|0\rangle\cr
  &=&\langle m(t/2)|m^*(t/2)\rangle\label{SE2}\;.
\end{eqnarray}
Equation~(\ref{SE2}) relates the overall spin echo magnetization
to the magnetization of the free induction decay.

\section{The Strong Collision Approximation and its Extension}

The analytical determination of the free induction decay according
to Eqs.~(\ref{Magn}) is restricted to very few cases, e.g. free
diffusion in a linear gradient or stochastic fluctuations between
two precession frequencies. The idea of the strong collision
approach and its extension is to replace the generator of the
Markov process $\R$ by a more simple generator $\D$ that conserves
specific features of the original dynamics.

\subsection{Strong Collision Approximation}

In many cases the stochastic fluctuations of the perturbation
fields occur on a much shorter time scale than spin-dephasing,
i.e. the correlation time $\tau$ of field fluctuations is much
shorter than the relaxation time of the magnetization. For ergodic
Markov processes one can estimate that after a few $\tau$ a spin
has visited almost all relevant states with the equilibrium
probability. On the other hand, there is only little change of the
magnetization during this time interval. Therefore, spin dephasing
in this situation can be described equivalently by a process in
which the transition rate between two states $i\to j$ is
independent of the initial state. Consequently, the transition
rate for $i\to j$ is proportional to the equilibrium probability
of the final state, $p_{0,j}$. Such a dynamics is referred to as
{\it strong collision dynamics}.

The generator $\D$ of this process has the form
\begin{equation}
\D=-\lambda\;({\bf id}-{\bf \Pi}_0)\label{strcollgen}\;
\end{equation}
where ${\bf \Pi}_0=|0\rangle\langle 0|$ is the projection operator
onto the eigenspace generated by the the equilibrium eigenvector
of $\R$, and ${\bf id}$ is the identity operator. The factor
$\lambda$ has to be determined self consistently.

Since the starting point of the strong collision approximation is
the observation that --- in many cases of interest --- the
correlation of the stochastic field fluctuations appear on a
shorter time scale than that of changes of the magnetization, only
the long time behavior of the field fluctuations is of importance.
This long time behavior is characterized by the correlation time
of the two-point autocorrelation function of the field
fluctuations (see also the appendix),
\begin{eqnarray}
C_2(t)&=& \langle\omega(t)\omega(0)\rangle\cr &=& \langle
0|\O\exp(\R t)\O|0\rangle \quad ,
\end{eqnarray}
which is defined as
\begin{eqnarray}
\tau_2&=&\int_0^{\infty}dt\;
\frac{C_2(t)-C_2(\infty)}{C_2(0)-C_2(\infty)}\cr\cr\cr
&=&\frac{\langle 0|\O\left[\exp(\R
t)-{\bf\Pi}_0\right]\O|0\rangle}{\langle
0|{\bf\Omega}^2|0\rangle-\langle
0|{\bf\Omega}|0\rangle^2}\label{2pointcf}
\end{eqnarray}
%%
%%%where $\hat{c}_2(s)$ is the Laplace transform of $c_2(t)$.
Stochastic field fluctuations determined by the strong collision
(SC) process should have the same correlation time as the original
process, leading to the self-consistency condition
\begin{equation}
\tau_2^{(SC)}(\lambda) = \tau_2\;. \label{CTSC}
\end{equation}
It is easy to determine that the correlation time for the strong
collision approximation is $\tau_2^{(SC)}(\lambda)=\lambda^{-1}$,
see the appendix Eq.~(\ref{SCtau}), leading to
\begin{equation}
\lambda=\tau_2^{-1} \;. \label{lambda}
\end{equation}

\subsection{Extended Strong Collision Approximation}

The extension of the strong collision approximation is based on a
comparison with the spectral expansion of the original operator,
\begin{equation}
\label{SpectralExpansion} {\bf
R}=\sum_{j=0}^{\infty}l_j\;{\bf\Pi}_j\;,
\end{equation}
where $l_0=0>l_1>\ldots$ are the ordered eigenvalues of $\bf R$
and ${\bf\Pi}_j=|j\rangle\langle j|$ is the projection operator
onto the eigenspace corresponding to $l_j$. Since the time
evolution operator is $\exp(\R\;t)=\sum_{j=0}^{\infty} e^{l_j
t}\;{\bf \Pi}_j$, it is clear that the low order eigenvalues
determine the long-time behavior, while higher orders dominate
shorter and shorter time scales. A comparison with a rewriting of
Eq.~(\ref{strcollgen}),
\begin{equation}
\D=l_0 {\bf \Pi}_0 -\lambda\;({\bf id}-{\bf \Pi}_0)\;
\label{strcollgenRewrite}
\end{equation}
(note that $l_0=0$), shows that in the strong collision
approximation just the lowest order term of
(\ref{SpectralExpansion}) is taken into account explicitly, while
the contribution of the higher eigenvalues is approximated by the
self-consistently determined parameter $\lambda$.

A natural extension, therefore, would be to take into account more
low order eigenvalues explicitly, thereby increasing the accuracy
of the description of the long-time behavior:
\begin{equation}
\label{gstrocollgen} {\D'}_n =
\sum_{j=1}^{n}l_j\;{\bf\Pi}_j\;-\lambda({\bf id}-{\bf\Pi})\;.
\end{equation}
with ${\bf\Pi}=\sum_{j=0}^n {\bf \Pi}_j$. A stochastic process
generated by an operator ${\D'}_n$ in Eq.~(\ref{gstrocollgen})
will be referred to as simplified extended strong collision
(ESC'$_n$) approximation of order $n$. As before, the contribution
of the higher eigenvalues is approximated  by the parameter
$\lambda$, which is determined again self-consistently from
condition (\ref{CTSC}). Here it leads to
\begin{equation}
\label{lambda2} \lambda = { c_2(0) -
\sum_{j=1}^{n}\left|\omega_{0i}\right|^2 \over
 c_2(0)\tau_2 + \sum_{j=1}^{n}l_j^{-1}\left|\omega_{0i}\right|^2 }
\quad , \label{lambda'}
\end{equation}
with $\omega_{0i}=\langle0|\O|i\rangle$ and
$c_2(0)=\langle0|\O^2|0\rangle-\omega_{00}^2$. Note that for $n\to
0$ this equation becomes (\ref{lambda}) again.

However, there are several problems involved with an approximation
based on (\ref{gstrocollgen}) and (\ref{lambda'}). Practically, an
exact determination of the low order eigenvalues and eigenvectors
is possible only in special cases. Therefore, one has to deal with
the problem that the eigenvalues and eigenvectors are known either
only approximately or not at all. Furthermore, even with
eigenvalues and eigenfunctions known, it turns out that the ESC'
approximation may be not applicable at all in certain situations:
If the autocorrelation function of the field fluctuations is
determined fully by the eigenfunctions included in ${\D}'$,
Eq.~(\ref{lambda'}) is undetermined. In that case additional
self-consistency requirements would be necessary for a better
description of the intermediate time regime.

Nevertheless, the above approach can be readily adapted to these
situations. Equation (\ref{gstrocollgen}) can be viewed as an
optimized reduced description of the relaxation in various
subspaces of the original operator $\R$. Such a optimized
description should also be possible for subspaces that are not
eigenvectors of $\R$. We can, therefore, set
\begin{equation}
\label{gstrocollgen2} {\D}_n =
-\sum_{j=1}^{n}\lambda_j\;{\bf\Pi}_j\;-\lambda({\bf
id}-{\bf\Pi})\;.
\end{equation}
However, now the rates $\lambda_j$, $j=1,\ldots,n$ are not
eigenvalues anymore, but have to be determined by additional
self-consistency requirements, see below. Moreover, the
${\bf\Pi}_j$ are not projectors onto the eigenspace of a
particular eigenvalue, but onto the spaces defined by arbitrarily
chosen mutually orthogonal functions $|f_j\rangle$,
$j=1,\ldots,n$, with $\langle f_i|f_{j}\rangle=\delta_{ij}$ and
$\langle f_j|0\rangle=0$; i.e. the projectors have the form
${\bf\Pi}_j=|f_j\rangle\langle f_j|$ and
${\bf\Pi}={\bf\Pi}_0+\sum_{j=1}^n {\bf \Pi}_j$. Naturally, one
would try to choose the functions $|f_j\rangle$ close to the
eigenfunctions $|j\rangle$, although it is not required for the
extension to work. Another natural function space for example is
based on
 polynomials in the frequency operator $\O$, i.e.
\begin{equation}
|f_i\rangle=p_i({\bf\Omega})|0\rangle\;\label{omegabase},
\end{equation}
where $p_i$ is some polynomial of degree $i$, the coefficients of
which are chosen in such a way that the orthonormal relations are
fulfilled. In the following we will refer to this base of
functions as the $\Omega$-base.

In analogy to Eq.~(\ref{gstrocollgen}) a stochastic process
generated by an operator ${\D}_n$ in Eq.~(\ref{gstrocollgen2})
will be referred to as extended strong collision (ESC$_n$)
approximation of order $n$.  It is evident that the ESC$_0$
approximation refers to the strong collision approximation.

We mentioned already that the rates $\lambda_j$, $j=1,\ldots,n$,
in Eq.~(\ref{gstrocollgen2}) have to be determined now by
additional self-consistency requirements. As it was with the SC
approximation, the aim of the ESC$_n$ approximation is to
approximate more closely the correlation of field fluctuations.
This is achieved by considering also higher order correlation
functions
\begin{eqnarray}
\lefteqn{C_m(t_{m-1},\ldots,t_1) = } \nonumber \\
 & & \quad \quad \left<\omega\left(
 {\scriptscriptstyle \sum_{j=1}^{m-1} } t_j\right)
\ldots\omega(t_2+t_1)\omega(t_1)\omega(0)\right> \nonumber\\
&=&\langle 0|{\bf\Omega}\exp({\bf
R}t_{m-1}){\bf\Omega}...\exp({\bf
R}t_1){\bf\Omega}|0\rangle\label{hocf} \;.
\end{eqnarray}
Following the same arguments as for the strong collision
approximation, the long time behavior of the $C_m$ is of interest.
In the same way as for the strong collision approximation this
should be characterized by some first order statistical moment
which is obtained by integration of the correlation function over
$t_{m-1}, ...,t_1$. However, direct usage of $C_m$ is hampered by
its non-vanishing asymptotic behavior: It is easily seen that from
the relation $\lim_{t_\nu\to\infty}\exp({\bf R }t_\nu)={\bf
\Pi}_0$ follows
\begin{eqnarray}
\lim_{t_\nu\to\infty} C_m(t_{m-1},..,t_1)&=&
C_{m-\nu}(t_{m-1},..,t_{\nu+1})\times\cr &&
C_\nu(t_{\nu-1},..,t_1) \quad ,
\end{eqnarray}
which does not necessarily vanish. In the strong collision
approximation we avoided this problem by considering the operator
$[\exp({\bf R}t)-{ \bf\Pi}_0]$ instead of $\exp({\bf R}t)$ in
Eq.~(\ref{2pointcf}), i.e. we considered only the relaxational
part of the stochastic process. When we perform the same
replacement in Eq.~(\ref{hocf}) we obtain modified correlation
functions $c_m(t_{m-1},\ldots,t_1)$ that we will call {\it quasi
cumulants} (s. appendix). They vanish asymptotically for all
$t_\nu$. We now require that the generalized correlation times
derived from these quasi cumulants,
\begin{equation}
\tau_{m}^{m-1} = \int_0^\infty \Pi_{i=1}^{m-1}\; dt_i
{c_m(t_{m-1},\ldots,t_1) \over c_m(0,\ldots,0)} \;,
\end{equation}
are equal for the exact process and for the extended strong
collision description. The relaxation rates are, therefore,
determined by
\begin{eqnarray}
\tau_m^{({ESC}_n)}(\lambda,\lambda_1,\ldots,\lambda_n) &=& \tau_m
\;, \nonumber \\
 & &  m = 2,4\ldots,2n+2 \;,
\label{CTSC2}
\end{eqnarray}
which replace the single self-consistency condition (\ref{CTSC}).
Note that in many systems the correlation functions
$c_m(t_{m-1},\ldots,t_1)$ vanish for odd values of $m$ due to
symmetry. Therefore, we require the equivalence of relaxation
times in Eq.~(\ref{CTSC2}) for even values of $m$ only. Otherwise
one has to determine the correlation times of the first $n+1$
non-vanishing correlation functions.

It is important to emphasize some properties of the ESC
approximation. First of all, it usually does not reduce to the
ESC' approximation when eigenfunctions are used for the projection
operator; i.e. the $\lambda_1,\ldots,\lambda_n$ do not take on the
numerical values of the corresponding eigenvalues, although they
usually do approximate them. In the light of the problems with the
ESC' approximations mentioned above, it will turn out that this is
an advantage: the self-consistent determination of the  relaxation
parameters $\lambda$, and $\lambda_1,\ldots,\lambda_n$ according
to Eq.~(\ref{CTSC2}) is more balanced than when only
Eq.~(\ref{CTSC}) is used, and gives rise to an improved
approximation. Moreover, the self-consistency conditions
(\ref{CTSC2}) imply that both processes, the ESC process and the
original Markov process have the same motional narrowing expansion
of the transverse relaxation, as it is shown in the
 appendix, Eq.~(\ref{mnarrowing2}).

\subsection{Transverse Spin Relaxation in the Extended Strong
Collision Approximation}

In this section we will exploit the simple structure of the
generator $\D_n$ to determine the time course of magnetization. We
will consider both: the free induction decay, i.e. the
superposition of coherent and incoherent spin dephasing, and the
spin echo decay, i.e. pure incoherent spin dephasing.

\subsubsection{Free Induction Decay}
In the extended strong collision approximation the generator $\R$
of the free induction decay in the generalized Bloch Torrey
Equation~(\ref{GBT}) is replaced by the generator $\D_n$ of
Eq.~(\ref{gstrocollgen2}). Instead of solving the propagator
$\U(t)=\exp((\D_n+{\rm i}\O)t)$ it is more convenient to solve its
Laplace transform ${\hat{\bf U}}(s)=(s-{\bf D}_n-{\rm i} \O)^{-1}$
which may be expanded as
\begin{equation}
\label{Laplace1} {\hat{\bf U}}(s)={\hat{\bf
U}}_0(s+\lambda)+{\hat{\bf U}}_0(s+\lambda)\;{\bf\Lambda}\;
{\hat{\bf U}}(s)
\end{equation}
where ${\hat{\bf U}}_0(s)=(s-{\rm i}\O)^{-1}$ is the Laplace
transform in the static dephasing limit (${\bf D}_n=0$), and the
operator ${\bf\Lambda}$ is defined as
\begin{equation}
{\bf\Lambda}=\sum_{j=0}^n(\lambda -\lambda_j)\; {\bf\Pi}_j\;,
\label{lambdamatrix}
\end{equation}
where we set $\lambda_0=0$. We will now confine the operators in
Eq.~(\ref{Laplace1}) onto the subspace defined by the projection
operator ${\bf\Pi}=\sum_{j=0}^n {\bf\Pi}_j$. Using the
abbreviation ${\bf O}^\Pi:={\bf\Pi}\;{\bf O}\;{\bf\Pi}$ for
denoting any operator ${\bf O}$ confined to that subspace, we
obtain
\begin{equation}
\label{Laplace2} {\hat{\bf U}}^\Pi(s)= {\hat{\bf
U}}_0^\Pi(s+\lambda)+{\hat{\bf
U}}_0^\Pi(s+\lambda)\;{\bf\Lambda}^\Pi\;{\hat{\bf U}}^\Pi(s)\;,
\end{equation}
where we exploited the fact that
${\bf\Lambda}={\bf\Pi}\;{\bf\Lambda}\;{\bf \Pi}$ and the
idempotency of projection operators, i.e. ${\bf\Pi}={\bf\Pi}^2$.
Equation~(\ref{Laplace2}) is of fundamental importance. It
demonstrates that the relation~(\ref{Laplace1}) between the
ESC-approximation and the static dephasing  is also valid in the
subspace $[\;|0\rangle,|f_1\rangle,..,|f_n\rangle\;]$. This
simplifies determination of spin relaxation considerably since one
only has to determine the $(n+1)\times (n+1)$ matrices
\footnote{in case of degeneracy of the eigenvalues the matrix
dimension is the sum of the dimensions of the eigenspaces plus
one} of the static dephasing limit ${\hat{\bf U}}_0^\Pi$ and
${\bf\Lambda}$, i.e.
\begin{equation}
\label{Laplace3}
 {\hat{\bf U}}^\Pi(s)=\big({\bf \Pi}-{\hat{\bf
U}}_0^\Pi(s+\lambda)\;{\bf\Lambda}^\Pi\big)^{-1}\;{\hat{\bf
U}}_0^\Pi(s+\lambda)\;.
\end{equation}
The Laplace transform of the overall magnetization decay
$\hat{M}_{[n]}(s)$ in the extended strong collision approximation
has the form
\begin{equation}
\label{propagatorESC} \hat{M}_{[n]}(s)=\langle
0|\;\hat{\U}^\Pi(s)\;|0\rangle
\end{equation}
For the special case of the strong collision approximation,
ESC$_0$, Eqs.~(\ref{Laplace3}) and (\ref{propagatorESC}) result in
\begin{equation}
\hat{M}_{[0]}(s)={{\hat{M}_{sd}
(s+\lambda)}\over{1-\hat{M}_{sd}(s+\lambda)\; \lambda}}\;.
\label{LSC}
\end{equation}
with $\hat{M}_{sd}(s)=\langle 0|{\hat{\bf
U}}_0(s+\lambda)|0\rangle$ as the Laplace transform of the overall
magnetization in the static dephasing regime. The time evolution
$M(t)$ can be obtained from Eqs~(\ref{Laplace3}) and
(\ref{propagatorESC}) either by numerical inverse Laplace
transform or using the generalized moment approach
\cite{bauer1,nadler} which allows a multi-exponential
approximation.

\subsubsection{Spin-Echo Decay}

The spin-echo decay is obtained by inserting of the generator
$\D_n$
%%%(Eq.~(\ref{gstrocollgen}))
into Eq.~(\ref{SE2}), i.e.
\begin{eqnarray}
\partial_t M_{SE,[n]}(t)&=&-\lambda\; M_{SE,[n]}(t)+\langle
m(t/2)|{\bf\Lambda}|m^*(t/2)\rangle\cr&=&-\lambda\;
M_{SE,[n]}(t)+\langle 0|{\bf U}(t/2){\bf\Lambda}{\bf
U}^*(t/2)|0\rangle\cr &=&-\lambda\; M_{SE,[n]}(t)+\langle 0|{\bf
U}^\Pi(t/2){\bf\Lambda}{\bf U}^{*\;\Pi}(t/2)|0\rangle\cr
%%&=&-\lambda\; M_{SE,[n]}(t)+\lambda\;\left|M_{[n]}
        %%(t/2)\right|^2\cr &&+
%%\sum_{i=1}^n(\lambda-\lambda_i)\langle 0\left| {\bf
        %% U}(t/2)|i\rangle \right|^2
%%% \langle i|{\bfU}^*(t/2)|0\rangle\cr
\label{ESC4}
\end{eqnarray}
i.e. the spin-echo decay is expressed as a function of the
spin-echo amplitude $M_{SE}$, and the projection of the free
induction decay onto the subspace $[\;|0\rangle,
|f_1\rangle,...,|f_n\rangle\;]$, i.e.  ${\bf U}^\Pi(t)|0\rangle$.
This projection of the free induction decay is obtained from
Eq.~(\ref{Laplace3})by inverse Laplace transform, i.e. ${\bf
U}^\Pi(t)|0\rangle={\cal L}^{-1}(\hat{\bf U}^\Pi(s)|0\rangle)$.
The solution of Eq.~(\ref{ESC4}) is
\begin{equation}
\label{ESC4solved} M_{SE,[n]}(t)=e^{-\lambda
t}\left[1+2\int_{0}^{t/2}d\xi e^{2\lambda\xi}\langle 0|{\bf
U}^\Pi(\xi){\bf\Lambda}{\bf U}^{*\Pi}(\xi)|0\rangle\right]
%\left[\lambda \;\left|M_{[n]} (\xi/2)\right|^2+ \right.\right.\cr
%&& \left.\left. \sum_{i=1}^n(\lambda-\lambda_i) \left|\langle
%0|{\bf U}(\xi/2)|i\rangle\right|^2
%%%\langlei|{\bf U}^*(\xi/2)|0\rangle \right]\right\}\cr
\end{equation}

\subsubsection{Time Constants of Transverse Relaxation}

The free induction and the spin echo decay are usually described
by the time constants $T_2^*$ and $T_2$. However, there is no
unique definition of these parameters. One definition of the
relaxation times is
\begin{eqnarray}
1/T_2^*&=&-\ln(M(t))/t \cr 1/T_2&=&-\ln(M_{SE})/t\;.\label{T2SE}
\end{eqnarray}
For the ESC decay one has to replace $M$ by $M_{[n]}$ and $M_{SE}$
by $M_{SE,[n]}$. This definition implies a dependence of
relaxation times on $t$, except for single exponential decay.

Another definition of relaxation times is based on the assumption
that these constants provide the best single exponential
approximation of magnetization decays, i.e. $M(t)\approx
e^{-t/T_2^*},\; M_{SE}\approx e^{-t/T_2}$. According to the mean
relaxation time approximation the relaxation times are then the
first long time moments of the decays \cite{nadler}, i.e.
\begin{eqnarray}
T_2^*:=\mu_{-1}(M)&=&\int_{0}^{\infty}dt\; M(t)\cr
T_2:=\mu_{-1}(M_{SE})&=&\int_{0}^{\infty}dt\; M_{SE}(t)\label{T2}
\end{eqnarray}
For a single exponential the mean relaxation time definition and
the definitions~(\ref{T2SE}) give the same results. According to
definition~(\ref{T2}) the relaxation times of the ESC decays can
be related to their Laplace transforms as
\begin{eqnarray}
T_2^*&=&\hat{M}_{[n]}(0)\cr
T_2&=&\hat{M}_{SE,[n]}(0)\;,\label{T2ESC}
\end{eqnarray}
The term $\hat{M}_{[n]}(0)$,  which provides $T_2^*$, is obtained
from Eq.~(\ref{propagatorESC}). Applying some rules of Laplace
transforms the term $\hat{M}_{SE,[n]}(0)$ giving $T_2$ is obtained
from Eq.~(\ref{ESC4})  as
%%
%\begin{eqnarray}
%\hat{M}_{SE,[n]}(s)&=&(s+\lambda)^{-1}\;(1+2\lambda\; {\cal
%L}\big( |M_{[n]}|^2,2s\big)\cr
%&&+2\;\sum_{i=1}^n(\lambda-\lambda_i)\;{\cal L}\big(\langle
%m|i\rangle\langle i|m^*\rangle,2s\big)\;.\label{ESC5}
%\end{eqnarray}
%The relaxation time $T_2$ of the spin echo decay is then
%%
\begin{equation}
T_2=\lambda^{-1}+2\;\sum_{i=0}^n(1-\lambda_i/\lambda)\;\Theta_i
\label{ESC6}\;,
\end{equation}
where
\begin{eqnarray}
\label{Theta0} \Theta_0&=&\int_{0}^{\infty}dt\; |\langle 0|{\bf
U}^\Pi(t)|0\rangle|^2\cr\cr &=&
\frac{1}{2\pi\i}\int_{-\i\infty}^{\i\infty}dz \;\langle
0|{\hat{\bf U}^\Pi}(z)|0\rangle\langle 0|{\widehat{\bf
U}^{*\Pi}}(-z)|0\rangle
\end{eqnarray}
is the mean relaxation time of the absolute squared overall free
induction magnetization $|M_{[n]}(t)|^2=|\langle 0|{\bf
U}(t)^\Pi|0\rangle|^2$, and for $i\ge 1$
\begin{eqnarray}
\label{Thetai} \Theta_i&=&\int_0^{\infty}dt\;\langle 0| {\bf
U}^\Pi(t)| f_i\rangle\langle f_i|{\bf
U}^{*\Pi}|0\rangle(t)\rangle\cr\cr &=&
\frac{1}{2\pi\i}\int_{-\i\infty}^{\i\infty}dz \;\langle
0|{\hat{\bf U}}^\Pi(z)|f_i\rangle\langle f_i|{\widehat{\bf
U}^{*\Pi}}(-z)|0\rangle
\end{eqnarray}
are transit times describing the transient occurrence of the
non-equilibrium components of the free induction decay ${\bf
U}(t)|f_i\rangle$. Equation~(\ref{ESC6}) relates $T_2$ which
describes the incoherent, i.e. irreversible, component of spin
dephasing to the stochastic field dynamics ($\lambda, \lambda_i$)
and time constants of the free induction decay ($\Theta_i$), i.e.
Eq.~(\ref{ESC6}) is a dissipation-fluctuation-coherence relation.
Note that the Eqs.~(\ref{Theta0}) to (\ref{Thetai}) directly
relate the time constants $\Theta_i$ to the Laplace transform of
the free induction decay $\hat{\bf U}^\Pi(s)$ given by the
fundamental Equation~(\ref{Laplace2}).

>From Eq.~(\ref{ESC6}) one can derive asymptotic relations for very
fast and slow stochastic field fluctuations. Let $\epsilon$ be
some scaling parameter of ${\bf D}_n$, i.e.
$\lambda,\;\lambda_i\sim\epsilon$, then Eq.~(\ref{ESC6}) reads in
the static dephasing limit ($\epsilon\to 0$)
\begin{equation}
T_2\approx \lambda^{-1}
\end{equation}
where we exploited that $\Theta_i(\epsilon)$ approaches its finite
static dephasing limit. For very fast fluctuations i.e. in the
motional narrowing limit ($\epsilon\to \infty$) one exploits that
$\langle f_i|{\hat{\bf U}}(s)|0\rangle/\langle 0|{\hat{\bf
U}}(s)|0\rangle\sim \epsilon^{-1}$, as a power expansion
demonstrates, i.e. one obtains
\begin{equation}
T_2\approx 2\;\Theta_0\; .
\end{equation}
This implies that the spin echo relaxation time is almost
identical with the relaxation time of the absolute squared
magnetization of the free induction decay, or vice versa that the
free induction decay is almost irreversible.

The dissipation-fluctuation-coherence relation (\ref{ESC6}) takes
a very simple form in the strong collision approximation, when we
assume that the overall magnetization decay is well approximated
by a single exponential, i.e. $M_{[0]}(t)\approx e^{-t/T2^*}$.
Since $\lambda=\tau_2^{-1}$, see Eq.~(\ref{lambda}),
Eq.~(\ref{ESC6}) reads
\begin{eqnarray}
T_2&=&\tau_2+2\Theta_0\cr&\approx& \tau_2+T_2^*\label{ESC7}\;.
\end{eqnarray}

>From Eqs.~(\ref{ESC7}) follows that in the motional narrowing
limit $T_2\approx T_2^*$ holds whereas in the static dephasing
limit of the strong collision approximation the relation
$T_2\approx \tau_2$ holds.

\section{Applications}

\subsection{Anderson-Weiss Model}

The Anderson-Weiss model \cite{anderson} is one of the rare
approaches -- besides the ESC approximation -- which describes
spin dephasing over the whole dynamic range of stochastic field
fluctuations. The approach is suitable, for example, when
dephasing is induced by spin interaction with a great number of
independently fluctuating perturbation fields in the spin
environment. Then analytical results are obtained for the free
induction and the spin echo magnetization decay as
\begin{eqnarray}
M(t)&=&\exp\bigg[-\int_0^t
(t-\xi)\;c_2(\xi)\;d\xi\bigg]\label{MFT2st}\\
M_{SE}(t)&=&\exp\bigg[-4\int_0^{t/2}(t/2-\xi)\;c_2(\xi)\; d\xi
\cr&&+\int_0^t(t-\xi)\; c_2(\xi)\; d\xi\bigg]\;, \label{MFT}
\end{eqnarray}
where $c_2$ is the modified two point correlation function (s.
appendix). In this section we will first characterize the class of
Markovian processes which fulfills the conditions of the
Anderson-Weiss model. This leads to a generalized Bloch-Torrey
equation according to Eq.~(\ref{GBT}) which is solved. Finally we
compare the Anderson-Weiss  model with its ESC$_0$ and ESC$_1$
approximation.

\subsubsection{Markovian and Anderson-Weiss Dynamics}

The Anderson-Weiss approach is based on a Gaussian distribution of
perturbation field frequencies $\omega$. Even more important is
the {\it additional} assumption that the stochastic phase
accumulation of a spin $\phi=\int_0^t d\xi\; \omega(\xi)$ also
exhibits a Gaussian distribution. This latter condition implies
that the Greens function $G(\omega_j,\omega_i,t)$, i.e. the
probability that a spin initially precessing with frequency
$\omega_i$ precesses at $t$ with $\omega_j$, is also a Gaussian
function in $\omega_j,\omega_i$ with the condition
$G(\omega_j,\omega_i,0)=\delta(\omega_j-\omega_i)$.
%% Markovian
%% processes have a Greens function of the form
%% $G(\omega_j,\omega_i,t)=\exp(\R t)_{j,i}$, where $\R$ is the
%% transition rate matrix. The Gaussian shape of the Greens function
%% implies that the transition rate from $\omega_i$ into the interval
%% $[\omega_j-\Delta,\omega_j+\Delta]$ vanishes, i.e.  $\lim_{t\to 0}
%% \big[t^{-1} \int_{\omega_j-\Delta}^{\omega_j+\Delta}
%% G(\xi,\omega_i,t) d\xi\big]=0$ as long as
%% $\Delta<|\omega_j-\omega_i|$.
This implies that only nearest neighbor transitions rates are
non-vanishing. Markovian processes in a continuous variable
$\omega$ with this property are described equivalently by a Fokker
Planck Equation \cite{Gardiner}, i.e. the probability density
$p(\omega)$ satisfies
\begin{eqnarray}
\partial_t p(\omega,t)&=&\R\;p(\omega,t)\cr &=&\partial_\omega\;
D(\omega)(\partial_\omega-F(\omega))\;p(\omega,t)\;,
\label{Fokker}
\end{eqnarray}
where $D(\omega)$ is a -- possibly $\omega$-dependent -- diffusion
coefficient and $F(\omega)$ is some driving force. Since the
equilibrium probability density is a Gaussian function one obtains
$F(\omega)=-c\cdot\omega$ with $c>0$. The generalized Bloch-Torrey
equation~(\ref{GBT}) which determines the dynamics of
magnetization as a superposition of precession and stochastic
transitions then reads
\begin{equation}
\partial_t m(\omega,t)=[\partial_\omega
D(\omega)(\partial_\omega+c\;\omega)+\i\;\omega]m(\omega,t)\;.
\label{FokkerBT}
\end{equation}
The derivation of the Eqs.~(\ref{Fokker}), (\ref{FokkerBT}) is of
fundamental importance since it states that a Markovian dynamics
of a variable $\omega$ which satisfies the Anderson-Weiss
conditions is equivalent to a diffusion process in this variable
within a harmonic potential $c\;\omega^2/2$ and vice versa.
Transformation of variables $\omega\to c^{1/2}\omega$ and $t\to
c^{-1/2} t$ simplifies Eq.~(\ref{FokkerBT}) to
\begin{equation}
\partial_t m(\omega,t)=[\partial_\omega\beta(\omega)
(\partial_\omega+\omega)+\i\;\omega] m(\omega,t)\label{nFokkerBT}
\end{equation}
where we continue to denote also the transformed variables as
$\omega$ and $t$ and $\beta=c^{3/2}\;D$ is the transformed
diffusion coefficient. In the following we will restrict
consideration to the case of a constant diffusion coefficient
$\beta$. The left and right sided eigenfunctions of the transition
operator $\R=\beta\partial_\omega(\partial_\omega+\omega)$ are the
Hermite functions, i.e.
\begin{eqnarray}
|n\rangle &\sim& \exp(-\omega^2/2) H_n(\omega) \cr \langle
n|&\sim& H_n(\omega) \label{Hermite}
\end{eqnarray}
with eigenvalues
\begin{equation}
l_n=-n\;\beta\quad .
\end{equation}

>From the definition of the Hermite functions and the operator
intertwining relation
$[\partial_\omega,(\partial_\omega+\omega)]=1$ follow the
recursive Equations
\begin{eqnarray}
&|n+1\rangle=-\frac{1}{n+1\;}\partial_\omega|n\rangle\;,\;
&|n-1\rangle=(\partial_\omega+\omega)|n\rangle\cr\cr &\langle
n+1|=\langle n|(\partial_\omega+\omega)\;,\; &\langle
n-1|=-\frac{1}{n}\;\langle n|\partial_\omega \label{rekursiv}
\end{eqnarray}
which also provide the normalization of eigenfunctions. The
advantage of the Markovian formulation of the Anderson-Weiss model
is that it does not only provide global parameters but also local
ones, e.g. the time course of the magnetization with frequency
$\omega$. Straightforward application of the Eqs.~(\ref{rekursiv})
and some operator algebra provides the solution of
Eq.~(\ref{nFokkerBT}) as
\begin{eqnarray}
m(\omega,t)&=&\exp[\beta\partial_\omega(\partial_\omega+\omega)+
\i\;\omega]| 0\rangle\cr\cr &=&\exp[-\beta^{-1}
t+\beta^{-2}(1-e^{-\beta t})]\cr
&&(2\pi)^{-1/2}\exp[-1/2(\omega-\i\;\beta^{-1}(e^{\beta t}-1))]
\label{magAW}
\end{eqnarray}
Integration over $\omega$ just gives the free induction decay of
the overall magnetization
\begin{equation}
M(t)=\exp[-\beta^{-1} t+\beta^{-2}(1-e^{-\beta t})] \label{MAW}
\end{equation}
which is just equivalent to the result of Eq.~(\ref{MFT2st}),
since the 2-point correlation function of Eq.~(\ref{nFokkerBT}) is
$c_2(t)=e^{-\beta t}$ (s. appendix Eq.~(\ref{corrf2ho})).
Insertion of this 2-point correlation function into
Eq.~(\ref{MFT}) provides the spin echo decay as
\begin{equation}
M_{SE}(t)=M(t/2)^2\;\exp[\beta^{-2}(e^{-\beta
t/2}-1)^2]\;.\label{MSEAW}
\end{equation}
Relaxation times of the free induction and spin echo decay were
determined according to Eqs.~(\ref{T2}).

\subsubsection{ESC approximation}
The ESC propagator is determined from the propagator in the static
dephasing limit $\U_0=\exp(\i\O t)=[\exp(\i\omega t)]$, and the
${\bf \Lambda}$-matrix of Eq.~(\ref{lambdamatrix}) both restricted
either to the function space $[\;|0\rangle\;]$ for the ESC$_0$ or
$[\;|0\rangle,\;|f_1\rangle\;]$ for the ESC$_1$ approximation
Eq.~(\ref{Laplace3}). The special structure of the transition rate
operator of the Anderson Weiss model implies that the base of
eigenfunctions, Eqs.~(\ref{Hermite}), is identical with the
$\Omega$-base, Eq.~(\ref{omegabase}), i.e.
$p_n(\O)|0\rangle=|n\rangle$. Hence, we will set in the following
$|f_1\rangle=p_1(\O)=|1\rangle$

\paragraph*{ESC$_0$ approximation}
The matrix element of the Laplace transformed propagator in the
static dephasing limit required for the ESC$_0$ approximation is
\begin{equation}
\label{FIDHOs} \langle
0|\hat{\U}_0(s)|0\rangle=\sqrt{\pi/2}\;e^{s^2/2}\;{\rm
cerf}(s/\sqrt{2})\quad,
\end{equation}
where ${\rm cerf}(z)=1-{\rm erf}(z)$ is the complementary error
function. The coefficient $\lambda$ which guarantees the self
consistency condition Eq.~(\ref{CTSC}) is determined form the
Eqs.~(\ref{SCtau},\ref{corrt1ho}) (s. appendix) as
\begin{equation}
\lambda=\beta\;,\label{SCMF1}
\end{equation}

\paragraph*{ESC$_1$ approximation}
The matrix elements of the Laplace transformed propagator in the
static dephasing limit required for the ESC$_1$ approximation in
the $\Omega$-base are that of Eq.~(\ref{FIDHOs}) and
\begin{eqnarray}
\langle 0|\hat{\U}_0(s)|1\rangle&=& N\;{\cal L}[\langle
0|\exp(\i\O t)\O|0\rangle]\cr &=& N \;(-\i){\cal
L}[\partial_t\langle 0|\exp(\i\O t)|0\rangle]\cr &=& N\; \i ( 1-s
\langle 0|{\hat U}_0(s)|0\rangle)\;, \label{FIDHO1}
\end{eqnarray}
where the factor $N$ generally is some normalization factor with
$N^2=\langle 0| \O^2|0\rangle$, i.e. in the case of the Anderson
Weiss model it is simply $N=1$. Consequently, using some
elementary rules of Laplace transforms, one derives the other
matrix elements as
\begin{eqnarray}
\langle 1|\hat{\U}_0(s)|0\rangle&=&\langle
0|\hat{\U}_0(s)|1\rangle\cr \langle
1|\hat{\U}_0(s)|1\rangle&=&N^2\; s\;(1-s\;\langle
0|\hat{\U}_0(s)|0\rangle)\;,\label{FIDHO}
\end{eqnarray}
It has to be stressed that the Equations (\ref{FIDHO1}) to
(\ref{FIDHO}) are generally valid for all ESC$_1$ approximations
in the $\Omega$-base.

The coefficients $\lambda,\lambda_1$ guaranteeing  the self
consistency condition Eq.~(\ref{CTSC2}) are obtained from
Eqs.~(\ref{HOT2},\ref{corrt1ho}) and
(\ref{tau4ESC1simple},\ref{corrt3ho})
\begin{eqnarray}
\lambda_1&=&\beta\cr
 \lambda&=&2\beta \label{SCMF2}
\end{eqnarray}

\paragraph*{Relaxation in the ESC$_0$ and ESC$_1$ approximation}
The matrix $\hat{\bf U}_0^\Pi(s)$ and the coefficients $\lambda$
and $\lambda_1$ determine the Laplace transformed ESC propagator
$\hat{\bf U}^\Pi(s)$ in Eq.~(\ref{Laplace3}), which itself is the
base for all other calculations. It directly provides $T_2^*$ when
defined as the first moment, Eq.~(\ref{T2ESC}), of the free
induction decay Eq.~(\ref{propagatorESC}). Insertion of $\hat{\bf
U}^\Pi(s)$ into Eqs.~(\ref{Theta0},\ref{Thetai}) provides
according to Eq.~(\ref{ESC6}) the relaxation time of the spin echo
decay when defined as its first long time moment
Eq.~(\ref{T2ESC}). Inverse Laplace transformation of $\hat{\bf
U}^\Pi(s)$ gives the ESC propagator ${\bf U}^\Pi(t)$ which itself
allows determination of the spin echo decay
Eq.~(\ref{ESC4solved}).

The relaxation time $T_2^*:=\mu_{-1}(M)$ of the Anderson-Weiss
process is well approximated by the ESC$_0$ and ESC$_1$
approximation over the whole dynamic range of stochastic field
fluctuations (Fig.~1). In the static dephasing regime all curves
approach $\lim_{\beta\to 0} \mu_{-1}^{-1}=\sqrt{2/\pi}$. The
successive approximation of the spin echo relaxation by the ESC
approximation is seen from the magnetization decay curves (Fig.~2)
and the curves showing the dependence of $T_2$ obtained by either
definition (Eqs.~(\ref{T2SE})-(\ref{T2})) on the diffusion
coefficient $\beta$ as Figure 3 demonstrates. The latter curves
all run parallel in the motional narrowing regime ($\tau_1(\langle
0|\O^2|0\rangle)^{1/2}=\beta^{-1}\ll 1$) and exhibit a similar
location of the maximum relaxation rate. Towards the static
dephasing regime ($\beta\to 0$) the rate of the Anderson-Weiss
process declines less than the rates of the ESC processes.

\subsection{Spin Dephasing by Restricted Diffusion in a Linear
Gradient Field}
\subsubsection{The Exact Process}
Whereas dephasing of free diffusing spins in a linear gradient
field can be treated analytically, only numerical solutions exists
for the restricted diffusion case \cite{duh}. On the one hand
restricted diffusion in a linear gradient field provides a simple
model to study principle features of spin dephasing by diffusion.
On the other hand treatment of this problem is not only of
academic interest as already mentioned in the Introduction. We
will approximate the free induction and spin echo decay of the
global magnetization for the case of restricted diffusion by the
strong collision approximation (ESC$_0$) and its first extension
(ESC$_1$). The ESC$_1$ approximation will be performed for both,
in the $\Omega$-polynomial base,  i.e. $|f_1\rangle\sim
{\bf\Omega}|0\rangle$ and in the eigenfunction base, i.e.
$|f_1\rangle=|1\rangle$.

We assume that the spins diffuse within an interval of size $L$ in
a linear gradient field $\omega(x)=g\;x$. Reflecting boundary
conditions at $x=\pm L/2$ imply that $\partial_x m(\pm
L/2,t)\equiv 0$. With $D$ as the diffusion coefficient and
$\R=D[\partial_x^2]$ the Bloch-Torrey Eq.~(\ref{GBT}) has the form
$\partial_t m(x,t)=(D[\partial_x^2]+{\rm i} g x) m(x,t)$, where
the brackets $[\; ]$ denote that application of the operator
$\partial_x^2$ is restricted to functions which fulfill the
reflecting boundary conditions. Transformation of variables $x\to
x/L$ and $t\to t\;g\;L$ results in
\begin{equation}
\partial_t m(x,t)=(\beta\;[\partial_x^2]+{\rm i}\;x)\;
m(x,t)\label{BTLG}\;,
\end{equation}
and vanishing derivatives at the edges of the unit interval
\begin{equation}
\partial_x m(\pm 1/2,t)\equiv 0\label{BC1}
\end{equation}
with the diffusion coefficient $\beta=D/(g L^3)$. We continue to
denote also the transformed variables as $x$ and $t$ to reduce the
number of symbols. When the initial magnetization is proportional
to the equilibrium probability, i.e. $m(x,0)\equiv 1$,  the
Laplace transform $\hat{m}(x,s)$ of the local magnetization decay
satisfies
\begin{equation}
(\beta\;[\partial_x^2]+{\rm i}\;x)\hat{m}(x,s)=-1\;.\label{BTLGL}
\end{equation}
The Equation~(\ref{BTLG}) was solved numerically. Integration of
the result over the unit interval provided the free induction
decay of the overall magnetization, and application of
Eq.~(\ref{SE2}) on the result gave the spin echo decay. When the
spin echo relaxation time was defined as the first statistical
moment of the magnetization decay Eq.~(\ref{T2}) was applied. For
determination of $T_2^*$, defined as the first moment of the free
induction decay, Eq.~(\ref{BTLGL}) was solved numerically, and
integration $\int_{-1/2}^{1/2}dx\;\hat{m}(x,s)=\hat{M}(s)$ gave
$T_2^*={\hat M}(0)=\mu_{-1}(M)$.

\subsubsection{ESC approximation}
The determination of the ESC$_0$ and ESC$_1$ approximation is
completely analogous to that for the Anderson Weiss model, except
that the $\Omega$-polynomial and the eigenfunction base are not
identical.

\paragraph*{ESC$_0$ approximation}
The equilibrium function for the restricted diffusion within the
unit interval is
\begin{equation}
|0\rangle=1\;
\end{equation}
i.e. one obtains
\begin{equation}
\langle 0|\hat\U_0(s)|0\rangle={\rm i} \;\ln\left(\frac{s-{\rm
i}/2}{s+{\rm i}/2}\right) \;.\label{LapFID1}
\end{equation}
The self consistency condition for the strong collision
approximation (\ref{CTSC}) determines the parameter $\lambda$ as
(s. appendix Eqs.~(\ref{SCtau}) (\ref{tau1lg}))
\begin{equation}
\label{SCC1} \lambda=10\;\beta
\end{equation}
Insertion of the results of Eqs.~(\ref{LapFID1}) and (\ref{SCC1})
into Eq.~(\ref{Laplace3}) determines the Laplace transformed
propagator in the ESC$_0$ approximation $\langle 0|\hat{\bf
U}(s)|0\rangle$ from which $T_2^*$, $T_2$ and spin-echo decay
curves are obtained.

\paragraph*{ESC$_1$ approximation in the $\Omega$-polynomial base:}
The lowest order function besides the equilibrium state in the
$\Omega$-polynomial base  has the form
\begin{eqnarray}
|f_1\rangle&=&\langle 0|\O|0\rangle^{-1/2}\;\O|0\rangle\cr
&=&2\sqrt{3}\; x
\end{eqnarray}
The matrix element (\ref{LapFID1}) and the Equations
(\ref{FIDHO1}) to (\ref{FIDHO}) then directly provide the static
dephasing operator $\hat{{\bf U}}_0(s)=(s- {\rm i}\; x)^{-1}$ in
the $[|0\rangle,|f_1\rangle]$ base. The parameters
$\lambda_1,\lambda$ of the ESC$_1$ approximation are determined
from the self consistency condition (\ref{CTSC2}), i.e. with
Eqs.~(\ref{HOT2}), (\ref{tau1lg})and   and
Eqs.~(\ref{tau4ESC1simple}),(\ref{tau3lg}) one obtains
\begin{eqnarray}
\lambda_1&=&10\;\beta\cr\cr
\lambda&=&\frac{443520}{8900}\beta\cr\cr&\approx&49.83\;\beta
\end{eqnarray}

\paragraph*{Development in the eigenfunction space:}
The normalized non equilibrium  eigenfunctions of the restricted
diffusion operator are
\begin{eqnarray}
|\nu\rangle&=&\sqrt{2}\;\sin(\nu\pi\;
x)\;\mbox{for}\;\nu=1,3,...\cr &=&\sqrt{2}\;\cos(\nu\pi\;
x)\;\mbox{for}\;\nu=2,4,...\; .\label{eigenfunctions1}
\end{eqnarray}
Since $[\partial_x^2]$ is a symmetric operator left and right
sided eigenfunctions are identical. With $|f_1\rangle=|1\rangle$
and $z=\pi(1/2+\i\; s)$ one obtains
\begin{eqnarray}
 \langle
0|\hat\U_0(s)|1\rangle&=&\sqrt{2}\big[\sinh(\pi s)\Ci(\xi)+\cr
&&\quad\quad \i\;\cosh(\pi
s)\Si(\xi)\big]\Big|_{\xi=-z^*}^{\xi=z}\cr \langle
1|\hat\U_0(s)|1\rangle&=&-2\arctan(2s)-\cr &&\quad
\big[\i\;\cosh(2\pi s)\Ci(\xi)+ \cr &&\quad\quad\quad \sinh(2\pi
s)\Si(\xi)\big]\Big|_{\xi=-2z^*}^{\xi=2z}\cr && \label{LapFID}
\end{eqnarray}
where $\Ci$ and $\Si$ denote the integral cosine and integral
sinus function respectively. The parameters $\lambda_1\;\lambda$
in the eigenfunction base are determined similarly as in the
$\Omega$-base (s. appendix) and one obtains
\begin{eqnarray}
\label{ESCAL}\lambda_1&\approx&9.89\;
\beta\cr\lambda&\approx&41.6\;\beta
\end{eqnarray}

\paragraph*{Relaxation in the ESC$_0$ and ESC$_1$ approximation}
Figure 4 demonstrates the first long time moment of the free
induction decay, that of the strong collision approximation, and
its first extension for both bases as a function of the diffusion
coefficient $\beta$. All curves show the same asymptotic behavior
in the static dephasing ($\beta \to 0$) and in the motional
narrowing limit ($\tau_1 \langle 0|x^2|0\rangle=1/120
\;\beta^{-1}\ll 1$). Furthermore the better approximation by the
ESC$_1$ curves compared to the ESC$_0$ curve in the intermediate
motion regime is evident. There is no significant difference
between the ESC$_1$ approximation in the eigenfunction and in the
$\Omega$-polynomial base.

The spin echo magnetization decay is shown in Fig.~5. Especially
in the long time behavior near the static dephasing regime, the
ESC$_1$ curves either in the
        eigenfunction space or in the
$\Omega$-space demonstrate a better approximation than the ESC$_0$
curve. This is also reflected by the dependence of spin echo
relaxation rate $1/T_2$ on the diffusion coefficient (Figs. 6,7).
When defined by the echo time (Eq.~(\ref{T2SE})) the ESC$_0$ and
ESC$_1$ curves run parallel with the curve obtained for restricted
the diffusion dynamics for short echo times. For longer echo times
and decreasing diffusion coefficients the ESC$_1$ curve provides a
better approximation. Again as for the free induction decay there
is no significant difference between ESC$_1$ approximations in the
eigenfunction and that in the $\Omega$-base.

\section{Summary and Discussion}

Analytical results on transverse spin relaxation due to stochastic
phase modulation exist mainly for limiting cases, like the
motional narrowing and the static regime. Perturbation approaches
are only valid close to their respective limits, and they diverge
as one tries to extend them towards the opposite motion regime.
Particularly the intermediate motion regime cannot be described
reliably by such a treatment.

We choose a different approach. Our aim was to approximate the
dynamics, assumed to be Markovian, by a more simple one that
conserves specific features of the original. The starting point
was the strong collision approximation\cite{bauer1} that assumes
the transition probability between two states being independent
from the initial state, an approximation that holds when spin
dephasing occurs on a time scale significantly longer than the
stochastic phase modulations. Hence, all states perpendicular to
the equilibrium state relax with the same exponential factor that
is determined self-consistently by comparison with the field
fluctuations.

Note that the motional narrowing limit as well as the static
dephasing regime are described correctly by this approximation.
Consequently, the error in the intermediate motion regime is
already less than it would be by perturbation approaches of a
comparable low order. Nevertheless, there is still room for
improvement. Also, one would like to have higher order
approximations that can be used to check the quality of low order
descriptions.

A systematic extension of the strong collision ansatz is to
include the relaxation of states of an appropriate finite function
base explicitly. We require that correlation times of original and
approximate dynamics are identical to a certain order. This self
consistency condition assures that both dynamics have the same
motional narrowing expansion of spin dephasing. As it was already
in the strong collision ansatz, spin dephasing is asymptotically
identical for both dynamics in the limit of the static motion
regime.

The finite function base of the ESC$_n$ approximation may be given
by the first $n$ ordered eigenfunctions of the generator of the
original phase modulations. Obviously, then the ESC generator
directly reflects the dynamics of original generator up to a time
scale corresponding to the $n$-th eigenvalue. For practical
applications the ESC approach within an eigenfunction space may be
a safe way to approximate spin dephasing. However, when the
determination of the eigenfunctions is tedious, the application of
the $\Omega$-base
$[|0\rangle,\;|f_1\rangle\sim{\bf\Omega}|0\rangle,
|f_2\rangle\sim{\bf\Omega}^2|0\rangle, \ldots]$ may be more
appropriate, at least for the ESC$_1$ approximation. Within the
$\Omega$-base the determination of the two- and four-point
correlation times (s. appendix) and the propagator in the static
motion regime, Eq.~(\ref{FIDHO}), is considerably simplified.

The mechanism by which the ESC$_1$ approach in the $\Omega$-base
works becomes evident by the following consideration: terms of the
motional narrowing expansion Eq.~(\ref{narrowing1}) may be
interpreted as repetitive interactions of the spin system with the
inhomogeneous field ${\bf \Omega}$ and intermediate evolution with
the free propagator  $\exp({\bf R}t_i)$. In the motional narrowing
limit, one obtains from Eq.~(\ref{mnarrowing2})
\begin{eqnarray}
1/T_2&=&{\hat c}_2(0)\cr\cr &=&\int_0^\infty dt\;\langle
0|{\bf\Omega}\exp({\bf R} t){\bf\Omega}|0\rangle\cr\cr &=&\langle
0|{\bf\Omega}^2|0\rangle\;\int_0^{\infty}dt\;\langle f_1|\exp({\bf
R} t)|f_1\rangle\;,\label{f1v}
\end{eqnarray}
where the factor $\langle 0|{\bf\Omega}^2|0\rangle$ is due to the
normalization of $|f_1\rangle$, $\langle f_1|f_1\rangle=1$.
Equation (\ref{f1v}) implies that in the motional narrowing limit
the long time behavior of spin dephasing solely depends on the
free propagator related relaxation of the state $|f_1\rangle$,
i.e. this state remains the only relevant one. Hence, it is
obvious that in the intermediate motion regime an ESC$_1$
approximation including the state $|f_1\rangle$ in its generator
is superior to the ESC$_0$ approximation.

Within the function base the propagator of spin dephasing is
directly related to the propagator of spin dephasing in the
absence of stochastic phase modulations. This specific feature of
the ESC dynamics tremendously facilitates the actual determination
of spin dephasing for the following reasons: ($i$) in many cases
the propagator in the static motion regime (which is an average
phase factor) may be determined analytically or at least
numerically; ($ii$) the determination of the propagator from that
in the static motion regime is self contained within the base,
i.e. it is obtained from a combination of finite dimensional
matrices.

The two lowest order ESC approximations were applied to two
generic models: spin dephasing in the Anderson Weiss model, i.e.
Gaussian frequency distribution and Gaussian transition dynamics,
and dephasing by restricted diffusion in a linear frequency
gradient. The reason for this choice was that -- besides their
generic character -- these models allow either an analytical
(Anderson Weiss) or, at least, a simple numerical treatment
(linear gradient) of magnetization decay. These features are
helpful to prove the ESC approach. For the Anderson Weiss model we
determined the corresponding Markov generator of the phase
modulations, which -- to our knowledge -- was done here for the
first time.

For both generic models the subsequent improvement by ESC$_n$
approximations of dephasing parameters and magnetization decays
could be demonstrated. One of our next aims will be the
application of the ESC approach to more realistic scenarios.

In closing, we would like to emphasize that the ESC approach is
actually not limited to spin dephasing only. It can be applied, in
principle, in any situation where the time behavior of complicated
observables of stochastic processes is of interest. In each case,
however, an appropriate function base has to be chosen,
corresponding to the $\Omega$-base for spin dephasing.

\section*{Acknowledgments}
We thank P.~Grassberger for a critical reading of the manuscript.
This work was supported by the Deutsche Forschungsgesellschaft:
Sonderforschungsbereich 355 ``Pathophysiologie der
Herzinsuffizienz'', SFB 237 `` Unordnung und gro{\ss}e
Fluktuationen'', and Graduiertenkolleg ``NMR'' HA 1232/8-1.

\begin{appendix}

\section{Autocorrelation Functions and Quasi-Cumulants}

The general $n$-point autocorrelation function of stochastically
fluctuating fields $\omega_j$  is defined as
\begin{eqnarray} &&C_n(t_{n-1}, ...,t_1):=\cr\cr
&&\sum_{j_{n-1},..,j_0}p(\omega_{j_{n-1}},
\sum_{i=1}^{n-1}t_i;...;\omega_{j_1},t_1;\omega_{j_0},0)\;
\prod^{n-1}_{\nu=0}\omega_{j_\nu}\;, \label{npointcorrfunction}
\end{eqnarray}
where $p(\omega_{j_{n-1}},
\sum_{i=1}^{n-1}t_i;...;\omega_{j_1},t_1;\omega_{j_0},0)$ is the
probability to find at $t=0$ the frequency $\omega_{j_0}$, at
$t=t_1$ the value $\omega_{j_1}$,... and at
$t=\sum_{i=1}^{n-1}t_i$ the frequency $\omega_{j_{n-1}}$.When the
stochastic dynamics is determined by a Markov process, this
probability can be factored into transition probabilities between
sequential states $i\rightarrow i+1$ after the interval $t_{i+1}$
and the initial ($t=0$) probability distribution, i.e.
\begin{eqnarray}
&&p(\omega_{j_{n-1}},\sum_{i=1}^{n-1} t_i ;...;\omega_{j_0},0)=\cr
&&\prod_{i=1}^{n-1} p(\omega_{j_i}\leftarrow\omega_{j_{i-1}},
t_{j_i}) \;p(\omega_{j_0},0)\;.
\end{eqnarray}
The transition probabilities after the interval $t_i$ are the
matrix elements of evolution operator $\exp(\R t_i)$. Since the
dynamics is assumed to be stationary the initial probability
$p(\omega_{j_0},0)$ is the equilibrium state probability
distribution, i.e. we can rewrite Eq.~(\ref{npointcorrfunction})
\begin{equation} C_n=\langle 0|\;\O\exp(\R t_{n-1})\O....
\exp(\R t_1)\O\;  |0\rangle\;, \label{npointcorrfunction2}
\end{equation}
where $\O=(\omega_j\delta_{j,k})$ is the diagonal frequency
matrix.  A modification of the correlation functions occurs if one
exchanges the evolution operator $\exp(\R t)$ with the operator
$\exp(\R t)-{\bf\Pi}_0$, where ${\bf \Pi}_0=|0\rangle\langle 0|$
is the projection operator onto the equilibrium state space. This
modified evolution operator describes the relaxation of
observables minus their equilibrium state values. The modified
autocorrelation functions will be denoted as quasi cumulants and
they are then defined as
\begin{eqnarray}
c_n=\langle 0|\;\O[&\exp&(\R t_{n-1})-{\bf \Pi}_0]\O....\cr
....[&\exp&(\R t_1)-{\bf \Pi}_0]\O\;|0\rangle\;.\label{corrf3}
\end{eqnarray}
The Laplace transform of the correlation function in
Eq.~(\ref{corrf3}) has the form
\begin{eqnarray}
\hat{c}_n(s_{n-1},..s_1)=\langle
0|\;\O&\left[{1\over{s_{n-1}-\R}}-{1\over{s_{n-1}}}\;{\bf
\Pi}_0\right]&\O....\cr\cr
....&\left[{1\over{s_1-\R}}-{1\over{s_1}}\;{\bf
\Pi}_0\right]&\O\;|0\rangle\;.\label{lapcorrf3}
\end{eqnarray}
This Laplace transform allows the determination of temporal
moments of the normalized autocorrelation function
$c_n(t_{n-1},..,t_1)/c_n(0,..,0)$ as the generalized correlation
times
\begin{equation}
\label{gcorrtimes} \tau_{n}^{n-1}=\hat{c}_n(0,..,0)/c_n(0,..,0)\;.
\end{equation}

\section{Motional Narrowing Expansion}
The motional narrowing expansion is a perturbation approach to
determine
 the
overall magnetization $M(t)$ -- or its Laplace transform
$\hat{M}(s)$ -- in terms of powers of the fluctuating fields $\O$.
It is based on the assumption, that the stochastic fluctuations
are more rapid than the precession frequencies of the perturbation
fields (motional narrowing limit). We will present a general
relation between the relaxation of the magnetization and the
correlation of the field fluctuations that contains the motional
narrowing limit as a limit case. The first step is to expand the
Laplace transform of the overall magnetization, Eq.~(\ref{Magn}),
in $\O$, i.e.
\begin{eqnarray}
\hat{M}(s)&=&\langle 0|\;{1\over{s-\R-{\rm i}\O}}\;|0\rangle\cr\cr
&=&\langle 0|\;(s-\R)^{-1}+ {\rm i}(s-\R)^{-1}\O(s-\R)^{-1}\cr
&&+{\rm
i}^2(s-\R)^{-1}\O(s-\R)^{-1}\O(s-\R)^{-1}+..|0\rangle\cr\cr
&=&s^{-1}+s^{-2}\;{\rm i}\;\langle 0| \O |0\rangle\cr
&&+s^{-2}\;{\rm i}^2\;\langle 0|\O(s-\R)^{-1}\O\rangle+...\cr\cr
&=&s^{-1}\bigg(1+s^{-1}\;\sum_{\nu=1}^{\infty}{\rm
i}^\nu\;\hat{C}_\nu(\underbrace{s,s,..,s}_{(\nu-1)\times})\bigg)\;,
\label{narrowing1}
\end{eqnarray}
where $\hat{C}_\nu$ are the Laplace transformed n-point
correlation functions of Eq.(\ref{npointcorrfunction2}). To avoid
singularities at $s=0$ it is better to consider $\hat{M}^{-1}(s)$.
When we set $q=s^{-1}\;\sum_{\nu=1}^{\infty}{\rm
i}^\nu\;\hat{C}_\nu(\underbrace{s,s,..,s}_{(\nu-1)\times})$ one
obtains
\begin{eqnarray}
\hat{M}^{-1}(s)&=& s\;\bigg(1+\sum_{\rho=1}^{\infty}\;
(-1)^\rho\;q^\rho\bigg)\cr\cr &=&s-\sum_{\nu=1}^{\infty}{\rm
i}^\nu\hat{C}_\nu\;+\; s^{-1}\sum_{\nu_1,\nu_2=1}^{\infty}{\rm
i}^{\nu_1+\nu_2} \hat{C}_{\nu_1}\hat{C}_{\nu_2}+...\cr ...+&&
(-1)^\rho\;s^{-(\rho-1)}\sum_{\nu_1,..,\nu_\rho=1}^{\infty}{\rm
i}^{\nu_1+..\nu_\rho} \prod_{m=1}^{\rho}\hat{C}_{\nu_m}\;+...\cr
\label{prerearrange}
\end{eqnarray}
Rearrangement of terms of equal order in $\O$ provides
\begin{equation}
\label{mnarrowing1} \hat{M}^{-1}(s)= s-\sum_{j=1}^{\infty}{\rm
i}^j\;K_j \;,
\end{equation}
where the coefficients $K_j$ have the form
\begin{eqnarray}
K_j&=&\hat{C}_j-s^{-1}\sum_{\nu_1+\nu_2=j}
\hat{C}_{\nu_1}\hat{C}_{\nu_2}+.. \cr
&&...+(-s)^{1-\rho}\sum_{\nu_1+..+\nu_\rho=j}\;\;
\prod_{m=1}^{\rho}\hat{C}_{ j_\rho}+...\cr\cr &&...+(-s)^{1-j}\;\;
\hat{C}_1^j
\end{eqnarray}
A comparison of this sum with the modified correlation functions
$c_j$, Eqs.~(\ref{corrf3}) and (\ref{lapcorrf3}), shows that
\begin{equation}
K_j=\hat{c}_j(\underbrace{s,...,s}_{(j-1)\times})\;,
\end{equation}
i.e. one obtains
\begin{equation}
\label{mnarrowing} \hat{M}^{-1}(s)= s-\sum_{j=1}^{\infty}{\rm
i}^j\; {\hat c}_j(\underbrace{s,s..,s}_{(j-1)\times})\;.
\end{equation}

Equation~(\ref{mnarrowing}) expands the relaxation of the
magnetization in terms of correlation functions to an arbitrary
order. The long time behavior of $M(t)$ is determined by the
Laplace transform in the range of small $s$, i.e. in this range
the relation
\begin{equation}
\label{mnarrowing2} \hat{M}^{-1}(s)\approx
s-\sum_{j=1}^{\infty}{\rm i}^j\; \underbrace{{\hat
c}(0,0..,0)}_{=\tau_j^{j-1}\;c_j(0,0..,0)}\;,
\end{equation}
is valid. The series in Eq.~(\ref{mnarrowing2}) contains terms of
magnitude $\le\langle 0|\O^\nu|0\rangle /l^\nu$, where $l$ denotes
non vanishing eigenvalues of $\R$. The latter determine the
fluctuation frequency. In the motional narrowing limit these
fluctuations are much higher than the precessing frequencies
$\langle 0|\O^\nu|0\rangle /l^\nu \ll 1$, i.e. after normalization
of the average perturbation field $\langle 0|\O|0\rangle$ to zero,
i.e. $c_1=0$, $M(t)$ is given a single exponential decay with the
well known result for the transverse relaxation rate as ${1\over
T_2}={\hat c}_2(0)=\tau_1\;\langle 0|\O^2|0\rangle\;.$

\section{Quasi Cumulants in the Strong Collision Approximation}
In the strong collision (ESC$_0$) approximation the quasi
cumulants take a very simple form. Insertion of the generator
$\D=-\lambda({\bf id}-\Pib)$ into Eq.~(\ref{lapcorrf3}) results in
\begin{equation}
\hat{c}_n^{(ESC_0)}(s_{n-1},..s_1)=
\prod_{i=1}^{n-1}\frac{1}{s_i+\lambda}\;
c_{n-1}(0,..,0)\;,\end{equation}
i.e. the quasi cumulant is a product of single exponential
functions $e^{-\lambda t_i}$ and the generalized correlation
times, Eq.~(\ref{gcorrtimes}), are all identical namely
\begin{equation}
\tau_n^{(SC)}=\lambda^{-1}\label{SCtau}\;.
\end{equation}

\section{Quasi Cumulants in the Extended Strong Collision
Approximation}

We restrict ourselves here to the $ESC_1$ Approximation and
determine the correlation functions and generalized relaxation
times for the $\Omega$-polynomial base only in order to show the
principle. Extensions to higher order approximations and to other
function bases are straightforward, although they may be more
tedious to calculate.

For $ESC_1$ the generator of the stochastic field fluctuations has
the form $\D=-\lambda_1\;\Pie-\lambda\;({\bf id}-\Pib-\Pie)$. We
will determine only the Laplace transforms of the 2- and 4-point
correlation functions since the 3-point correlation functions
vanishes in the models we consider. According to
Eq.~(\ref{lapcorrf3}) the determination of the correlation
functions requires of the operator
\begin{equation}
\frac{1}{s-{\bf D}}-\frac{1}{s}{\bf\Pi}_0=
\frac{1}{s+\lambda_1}{\bf\Pi}_1+\frac{1}{s+\lambda}({\bf
id}-{\bf\Pi}_0-{\bf\Pi}_1)
\end{equation}
Assuming that the average frequency vanishes, i.e.
$\big<\O\big>=\langle 0|\O|0\rangle=0$, which can always be
achieved by normalization., $|f_1\rangle\sim\O|0\rangle$. Hence,
the projector ${\bf\Pi}_1=|f_1\rangle\langle f_1|$ takes the form
\begin{equation}
{\bf\Pi}_1=\frac{{\bf\Omega}|0\rangle\langle
0|{\bf\Omega}}{\langle 0|{\bf\Omega}^2|0\rangle}\;.
\end{equation}
For the Laplace transformed two point correlation function one
obtains then
\begin{equation}
\hat{c}_2^{(ESC_1)}(s)=(\lambda_1+s)^{-1}\;\langle
0|{\bf\Omega}^2|0\rangle\quad \label{HOC2},
\end{equation}
i.e. the two point correlation function exhibits a single
exponential decay with relaxation rate
\begin{equation}
\tau_2^{(ESC_1)}=\lambda_1^{-1}\quad \label{HOT2}.
\end{equation}
The 4-point correlation function is
\begin{eqnarray}
\lefteqn{\hat{c}_4^{(ESC_1)}(s_3,s_2,s_1)= } \nonumber \\
 & & \quad
\frac{1}{(s_3+\lambda_1)(s_1+\lambda_1)}
\bigg(\frac{1}{s_2+\lambda_1}-
\frac{1}{s_2+\lambda}\bigg)\frac{\langle
0|{\bf\Omega}^3|0\rangle^2}{\langle 0|{\bf\Omega}^2|0\rangle}
\cr\cr &+&\frac{1}{(s_3+\lambda_1)(s_2+\lambda)
(s_1+\lambda_1)}\bigg(\langle 0|{\bf\Omega}^4|0\rangle-\langle
0|{\bf\Omega}^2|0\rangle^2\bigg)\cr \label{HOC4}\end{eqnarray}
The four point correlation time is then determined as
\begin{eqnarray}
\tau_4^{(ESC_1)}&=&\bigg[\hat{c}_4^{(ESC_1)}(0)/c_4^{(ESC_1)}
(0)\bigg]^{1/3} \cr\cr
&=&\bigg[\frac{1}{\lambda_1^2}\bigg(\frac{1}{\lambda_1}-
\frac{1}{\lambda}\bigg)\frac{\langle
0|{\bf\Omega}^3|0\rangle^2}{\langle 0|{\bf\Omega}^2|0\rangle
(\langle 0|{\bf\Omega}^4|0\rangle-\langle
0|{\bf\Omega}^2|0\rangle^2)}\cr\cr
&&+\frac{1}{\lambda_1^2\lambda}\bigg]^{1/3}\cr\label{tau4ESC1}
\end{eqnarray}
In the case of the Anderson Weiss model and for the restricted
diffusion linear gradient one has $\langle
0|{\bf\Omega}^3|0\rangle=0$, i.e. Eq.~(\ref{tau4ESC1}) simplifies
to
\begin{equation}
\tau_4^{(ESC_1)}=\sqrt[3]{\frac{1}{\lambda_1^2\lambda}}
\label{tau4ESC1simple}
\end{equation}

\section{Relaxation Times of Correlation Functions in the Models}

In this final appendix  we will determine the generalized
relaxation times of stochastic field fluctuations up to the 4-th
order for the generic models we discuss in the main text.

\subsection{Diffusion in a Harmonic Potential}

When field fluctuations result from diffusion in a harmonic
potential according to Eq.~(\ref{Fokker}) and spin dephasing is
described by Eq.~(\ref{nFokkerBT}), the corresponding Laplace
transformed 2-point correlation function, Eq.~(\ref{lapcorrf3}),
is
\begin{eqnarray}
\hat{c}_2(s)&=&\langle0| \omega\frac{1}{s-\beta\partial_\omega
(\partial_\omega+\omega)}\omega|0 \rangle\cr &=&\langle
1|\frac{1}{s-\beta\partial_\omega
(\partial_\omega+\omega)}|1\rangle\cr &=&\frac{1}{s+\beta}\;,
\label{corrf2ho}
\end{eqnarray}
where we applied the operator properties of $\partial_\omega,\;
(\partial\omega+\omega)$ according to Eqs.~(\ref{rekursiv}). This
result shows that the two point correlation function exhibits a
single exponential decay.  Since $c_2(0)=\langle
0|\omega^2|0\rangle=1$ one obtains
\begin{equation}
\tau_2=\beta^{-1} \label{corrt1ho}
\end{equation}

Since $c_3(t)$ vanishes, the next relevant correlation function is
$c_4(t)$. Similarly, one obtains for its  Laplace transform
\begin{equation}
\hat{c}_4(s_3,s_2,s_1)=\frac{2}{(s_3+\beta)
(s_2+2\beta)(s_1+\beta)}\;. \label{corrf4ho}
\end{equation}
And since $c_4(0)=\langle 0|\omega^4|0\rangle=2$ the corresponding
correlation time is
\begin{equation}
\tau_4=\sqrt[3]{\frac{1}{2}}\;\beta^{-1} \label{corrt3ho}
\end{equation}

\subsection{Restricted Diffusion in a Linear Gradient Field}

In this section we will determine the correlation times $\tau_n$
of Eq.~(\ref{gcorrtimes}) for $n=2,\;4$ for 1-d restricted
diffusion of spins in a unit box in which they are affected by a
linear gradient field. In dimensionless parameters one obtains for
the generator $\R=\beta\;[\partial_x^2]$ where $\beta$ is the
dimensionless diffusion coefficient. The brackets denote that the
application of the operator $\partial_x^2$ is restricted to
functions with vanishing derivative at $x=\pm 1/2$ (reflecting
boundary conditions). The frequency operator is $\O=x$ and the
equilibrium state eigenfunction is $|0\rangle\equiv 1$.
%% The other
%% normalized eigenfunction of $[\partial_x^2]$ are
%% %%
%% \begin{eqnarray}
%% |\nu\rangle&=&\sqrt{2}\;\cos(\nu x)\;\mbox{for}\;\nu=2,4,...\cr
%% &=&\sqrt{2}\;\sin(\nu
%% x)\;\mbox{for}\;\nu=1,3,...\label{eigenfunctions}
%% \end{eqnarray}
%% %%
%% Since the diffusion operator is symmetric, the left eigenfunctions
%% $\langle\nu|$ are identical to the right eigenfunctions
%% $\nu\rangle$.

The determination of  correlation times $\tau_n $ requires the
calculation of the Laplace transformed quasi cumulants
$\hat{c}_n(0,..,0)$, Eq.~(\ref{gcorrtimes}). The definition of
this quasi cumulants according to Eq.~(\ref{lapcorrf3}) shows the
need of recurrent determination of terms of the form
\begin{equation}
f=\lim_{s\to 0}[-(s-\R)^{-1}+s^{-1} \Pib] g\;,\label{HE1}
\end{equation}
with some function $g$, i.e. $f$ fulfills
\begin{equation}
\label{HE2} {\bf R} f=(1-\Pib)\; g\;,
\end{equation}
i.e. with ${\bf R}= [\partial_x^2] $ Eq.~(\ref{HE2}) becomes a
second order differential equation. The application of the inverse
second order differential operator leaves in general 2 integration
constants. One may be determined from the reflective symmetric
boundary conditions; however, a further condition is needed to get
the second constant. A spectral decomposition $\R$ in
Eq.~(\ref{HE1}) shows $\Pi_0 f\equiv 0$, i.e. we obtain as a
further condition
\begin{equation}
\int_{-1/2}^{1/2} dx\;f(x)=0\;\label{vanishf}.
\end{equation}
\paragraph*{Two point correlation time:}
We define
\begin{eqnarray}
f_1(x)&=&\lim_{s\to 0}(-(s-\R)^{-1}+s^{-1}\Pib)\;x\;|0\rangle\cr
&=&\beta^{-1}\;(x^3/6-x/8)\label{f1}
\end{eqnarray}
which obviously fulfills the reflecting boundary condition at
$x=\pm 1/2$ and Eq.~(\ref{vanishf}). Hence, the two point Laplace
transformed quasi cumulant is
\begin{eqnarray}
\hat{c}_2(0)&=&-\langle 0|\;x\;\lim_{so
0}(-(s-\R)^{-1}+s^{-1}\Pib)\;x\;|0\rangle\cr
&=&\int_{-1/2}^{1/2}dx\;xf_1(x)=\beta^{-1}\frac{1}{120}\;
\end{eqnarray}
and with
\begin{equation}
c_2(0)=\langle 0|x^2|0\rangle =\int_{-1/2}^{1/2}dx\;x^2=1/12
\end{equation}
one obtains
\begin{equation}
\tau_2=\frac{1}{10}\;\beta^{-1}\label{tau1lg}\;.
\end{equation}

\paragraph*{Four point correlation time: }
Iterative application of Eq.~(\ref{HE1}) defines
\begin{equation}
f_2(x)=\lim_{s\to 0}(-(s-\R)^{-1}+s^{-1}\Pib)\;xf_1(x) \quad ,
\end{equation}
which, according to Eq.~(\ref{HE2}), fulfills
\begin{eqnarray}
[\partial_x^2] f_2(x)&=&(1-\Pib)\; xf_1(x)\cr
&=&xf_1(x)-\int_{-1/2}^{1/2}dx\;xf_1(x)\cr
&=&xf_1(x)+\frac{1}{120}\beta^{-1}
\end{eqnarray}
Insertion of $f_1$, Eq.~(\ref{f1}), and considering the reflective
boundary conditions and the condition (\ref{vanishf}) yields
\begin{equation}
f_2(x)=\beta^{-2}\left(\frac{x^6}
{180}-\frac{x^4}{96}+\frac{x^2}{240}-\frac{ 37}{161280}\right)\;.
\end{equation}
Similarly to the procedure above we could determine a function
$f_3(x)$ but instead we use a different approach which exploits
the symmetry of eigenfunctions of the operator
$\R=\beta[\partial_x^2]$. The Laplace transformed 4-point quasi
cumulant  is
\begin{eqnarray}
\hat{c}_4(0,0,0)&=&-\overbrace{\langle 0|\;x\; \lim_{s_3\to
0}\left[ -{1\over{s_3-\R}}+{1\over{s_3}}\;\Pib\right]}^{=f_1(x)}\;
x \cr &&
  \lim_{s_2\to
0}\left[-{1\over{s_2-\R}}+{1\over{s_2}}\;\bf\Pib\right]\cr &&x
\lim_{s_1\to 0}\left[-{1\over{s_1-\R}}+{1\over{s_1}}\;
\bf\Pib\right]x|0\rangle\cr\cr
&=&-\;\int_{-1/2}^{1/2}dx\;f_1(x)xf_2(x)\cr
&=&\beta^{-3}\frac{89}{79833600}\;,
\end{eqnarray}
and with
\begin{equation}
c_4(0,0,0)=\langle 0|x^4|0\rangle-\langle
0|x^2|0\rangle^2=\frac{1}{180}\;,\label{c4restdiff}
\end{equation}
we finally have
\begin{equation}
\label{tau3lg} \tau_4=\sqrt[3]{\frac{89}{443520}}\;\beta^{-1}
\end{equation}

\end{appendix}

\begin{figure}
\caption{Relaxation time $T_2^*$ (defined as the first long time
moment $\mu_{-1}$) of the free induction decay in the Anderson
Weiss model (AW) and its ESC$_0$ and ESC$_1$ approximation as a
function of the diffusion coefficient $\beta$. \label{fig:fig1}}
\end{figure}

\begin{figure}
%\vskip 1truecm \epsfig{file=Fig2.eps,width=8.5truecm} \vskip
%1truecm
\caption{ Spin-echo magnetization decay in the Anderson Weiss
model (AW) and its ESC$_0$ and ESC$_1$ approximations for three
different diffusion coefficients $\beta$. Note: for the diffusion
coefficient close to the motional narrowing regime
($\beta=10^{0.5}$) the Anderson Weiss and the ESC curves almost
run parallel. Therefore for the clearness of the figure, only the
Anderson Weiss curve is shown. In the intermediate motion regime
$\beta=1$ the original and the ESC$_1$ curve still run parallel
whereas the ESC$_0$ approximation already shows a moderate
deviation. Towards the static dephasing regime ($\beta=10^{-0.5}$)
the successive improved approximation of the Anderson Weiss curve
by the ESC curves is evident. \label{fig:fig2} }
\end{figure}

\begin{figure}
\caption{Dependence of the spin echo relaxation time $T_2$ on the
diffusion coefficient $\beta$ for the Anderson-Weiss model (AW)
and its ESC$_0$ and ESC$_1$ approximation: (a) The relaxation time
was defined according to Eq.~(\ref{T2SE}) by the echo time $t$,
and (b) as the first long time moment $\mu_{-1}$ of spin-echo
magnetization decay according to Eq.~(\ref{T2}). The Anderson
Weiss curves and the corresponding approximations converge as
$\beta$ approaches the motional narrowing regime $\tau_1(\langle
0|\O^2|0\rangle)^{1/2}=\beta^{-1}\ll 1$.  When defined by the echo
time (a) the $T_2$ curves of the Anderson Weiss model and its
approximations all run parallel for the short echo time ($t=1$).
With increasing echo time ($t=3,\;6$) the successive ESC
approximation becomes evident. \label{fig:fig3}}
\end{figure}

\begin{figure}
%\vskip -2.5truecm
%\epsfig{file=Abb1.EPS,width=8.5truecm} \vskip 1truecm
\caption{Relaxation time of the free induction decay $T_2^*$ of
spins diffusing within a linear field gradient in the unit
interval as a function of the diffusion coefficient $\beta$.
$T_2^*$ is defined as the first long time moment $\mu_{-1}$ and
obtained from the Bloch Torrey (BT) Equation (\ref{BTLG}). The ESC
approximations are shown. The ESC$_1$ approximation was determined
for the eigenfunction (ef) and the $\Omega$-base.
\label{fig:fig4}}
\end{figure}

\begin{figure}
%\vskip 1truecm \epsfig{file=Fig2.eps,width=8.5truecm} \vskip
%1truecm
\caption{Spin-echo magnetization decay for restricted diffusion
within a linear field gradient in the unit interval as obtained
from the Bloch Torrey (BT) Equation (\ref{BTLG}). Three diffusion
coefficients $\beta$ are considered. The ESC approximations in the
different diffusion regimes are demonstrated. The ESC$_1$
approximation was obtained for the eigenfunction (ef) and the
$\Omega$-base. \label{fig:fig5} }
\end{figure}

\begin{figure}
%\vskip 1truecm \epsfig{file=Abb3.eps,width=8.5truecm} \vskip
%1truecm
\caption{Spin echo relaxation time $T2$ as a function of the
diffusion coefficient $\beta$ for restricted diffusion in the unit
interval and the corresponding ESC approximations. The labeling of
the curves is as in Figure 4. The relaxation time was defined by
the echo time $t$ according to Eq.~(\ref{T2SE}). \label{fig:fig6}
}
\end{figure}

\begin{figure}
%\vskip 1truecm \epsfig{file=Abb3.eps,width=8.5truecm} \vskip
%1truecm
\caption{Spin echo relaxation time $T2$ as a function of the
diffusion coefficient $\beta$ for restricted diffusion in the unit
interval and the ESC approximations. Labeling is as in Figure 4.
The relaxation time is defined as the first long time moment
$\mu_{-1}$ of the spin echo decay Eq.~(\ref{T2}).
\label{fig:fig7}}
\end{figure}

\end{multicols}

\end{document}